\documentclass[preprint2]{aastex}
\def\beq{\begin{equation}}
\def\eeqno#1{\label{#1}\end{equation}}
\def\sss{\scriptscriptstyle}
\def\^#1{^{\sss #1}}
\def\_#1{_{\sss #1}}

\def\az{a\_0}
\begin{document}
\title{Light and Dark in the Universe}
\author{Mordehai Milgrom} \affil{DPPA, Weizmann Institute of
Science, Rehovot 76100, Israel}

\begin{abstract}
Translation of a popular article on MOND vs. dark matter that appeared, in Hebrew, in the Israeli magazine {\it Odyssea}, dedicated to ``thoughts and ideas in the forefront of science and philosophy''.
\end{abstract}
\maketitle

\section{Introduction}
A bizar world picture emerges when we interpret observations of our universe by the theories of Newton and Einstein. According to this picture--embraced by most physicists--only a small fraction of the material content of the universe is made of ``luminous'' or ``visible'' matter: the kind of matter familiar to us all, which can be ``seen'' using various astronomical auxiliaries, of which are made, for instance, stars, gas clouds, and planets. The standard, accepted wisdom, anchored in Newtonian and Einsteinian dynamics, forces us to invoke two additional ``dark'' components, which make up the lion share: ``dark matter'' and ``dark energy''. These are invisible components that differ from each other substantially, but the nature of either is still a mystery.
\par
In opposition, a cohort of astrophysicists and physicists, with me in their number, hold the view that there is no need for ``dark'' components in the universe; the ``visible matter'' suffices. However, interpreting the observations requires the introduction of a new theory of dynamics, which I proposed some thirty years ago under the name of ``MOND'' (for ``modified Newtonian dynamics''). This theory differs materially from those of Newton and Einstein, especially when applied to cosmological phenomena: from individual galaxies to the universe at large.
\par
These two paradigm are locked in a mortal combat, with most astrophysicists putting on MOND the odds David had been given against Goliath.
\par
The laws of dynamics embody the laws of gravity and of inertia. They describe, e.g., how a body moves under the gravitational influence of other bodies in a given system: For example, how planets move in aggregates like the solar system, how stars and gas clouds move under the gravitational influence of a galaxy they reside in, how galaxies move in clusters of galaxies, and indeed, how the universe, as a whole, develops under its own gravity. Newton's dynamics ruled supreme for abut 250 years, until the beginning of the 20th century. Then, in two separate moves, the theory of dynamics was improved and changed unrecognizably. 
\par
In one move, led mainly by Einstein, Newton's theory was replaced by relativity, first the special, then the general. Einstein's dynamics departs greatly from Newton's in phenomena involving motions close to the speed of light, and when gravitational energies for a unit mass are high.
\par
In another move, started by Max Planck, and later led by many physicists, Newtonian dynamics was replaced by quantum theory, which holds sway primarily for microscopic phenomena. Despite much effort by many researchers we do not yet have a full fledged and complete theory, one that would replace Newton's theory in describing phenomena in which both departures from this theory are important. This fact always reminds us that we are a far cry from a complete understanding of our universe.
\par
MOND proposes to modify dynamics in yet another domain of phenomena, not accessible to physicists in earlier generations.
\section{The puzzle of the mass discrepancy}
When interpreting the observations in astrophysics and cosmology, one employs Newtonian dynamics and Einsteinian dynamics, according to the problem at hand (quantum theory is not relevant for the description of the phenomena we deal with here).  The relevant analysis of galaxy dynamics is generally performed in two steps (a more detailed example will be given below): First, one uses the laws of dynamics to calculate the velocities of stars and gas clouds in a given galaxy. The calculation is based on the gravity of only the visible matter--visible by whatever astronomical tool is available, such as optical, radio, or x-ray telescopes. This is the only kind of matter whose existence is certain. It is made up of electrons, proton, atoms, and other known particles. The calculated velocities are such that they have to balance gravity exactly, so that constituents of a galaxy would neither escape from it, if the velocities are too high, nor collapse, if they are too small.
\par
In the second stage of the test, after these balancing speeds are calculated, they are compared with the speeds that are actually measured in the galaxy under study. Then, a striking fact is revealed: The standard dynamics fail this test miserably: the calculated speeds fall much shorter of the measured ones. In other words, according to the standard dynamics, the gravity provided by the visible matter is much too weak to arrest the escape of most parts of the  galaxy from escaping and dispersing into space, hence leading to the quick dismantling of the galaxy. This is unacceptable since galaxies are known to be long lived.
This conundrum appears in all galaxies, in cluster of galaxies, in super clusters, and even when the evolution of the universe at large is considered.
This glaring discrepancy between predictions and measurements is called ``the mass discrepancy'': the mass of the visible matter is much too small to balance the inertial forces indicated by the measured speeds.
\begin{figure}
\begin{center}
\includegraphics[width=0.9\columnwidth]{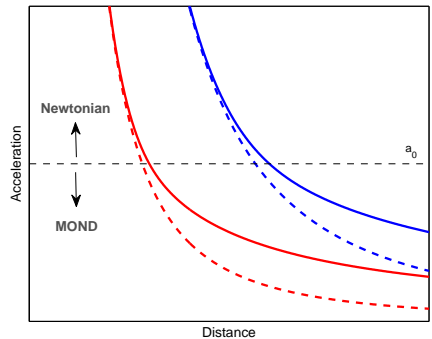}
\caption{The acceleration (the force acting on a unit mass) felt by a small body (e.g., a planet), as a function of its distance from a concentrated, massive body (e.g., a star like the sun), according to Newton's laws (dashed) and according to MOND (continuous). The blue lines correspond to an attractive mass four times more massive than in the red lines. We see that MOND departs from Newtonian laws always below the same acceleration, $\az$; this happens at different distances for different masses. We also see that the MOND acceleration never fall below the Newtonian one, for a given mass.}\label{fig1}
\end{center}
\end{figure}

\begin{figure}
\begin{center}
\includegraphics[width=0.9\columnwidth]{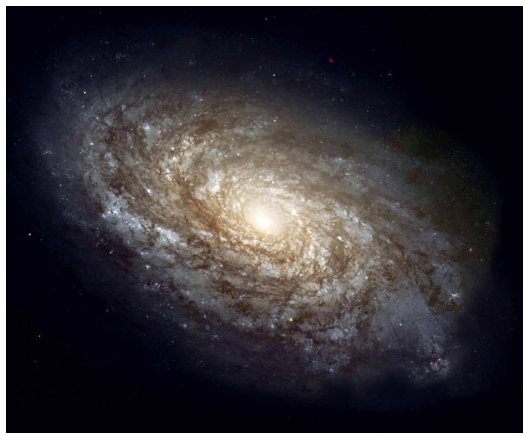}
\caption{A disc galaxy, such as those for which one measures rotation curves, from which the most clear-cut evidence for MOND emerges.}\label{fig2}
\end{center}
\end{figure}

\begin{figure}
\begin{center}
\includegraphics[width=0.9\columnwidth]{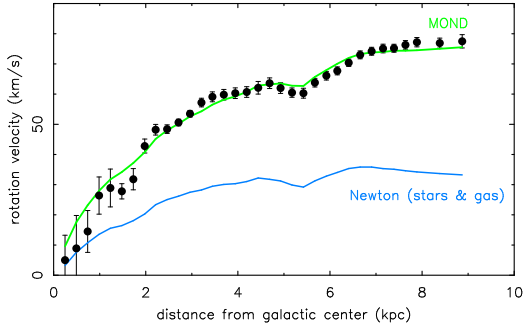}
\caption{The rotation curve measured for the disc galaxy NGC 1560 (the data points). The curve predicted by Newtonian dynamics, based on the luminous mass distribution, is shown by the blue line, and that predicted by MOND from the same mass distribution by the green line. Courtesy of Stacy McGaugh.}\label{fig3}
\end{center}
\end{figure}

\begin{figure}
\begin{center}
\includegraphics[width=0.9\columnwidth]{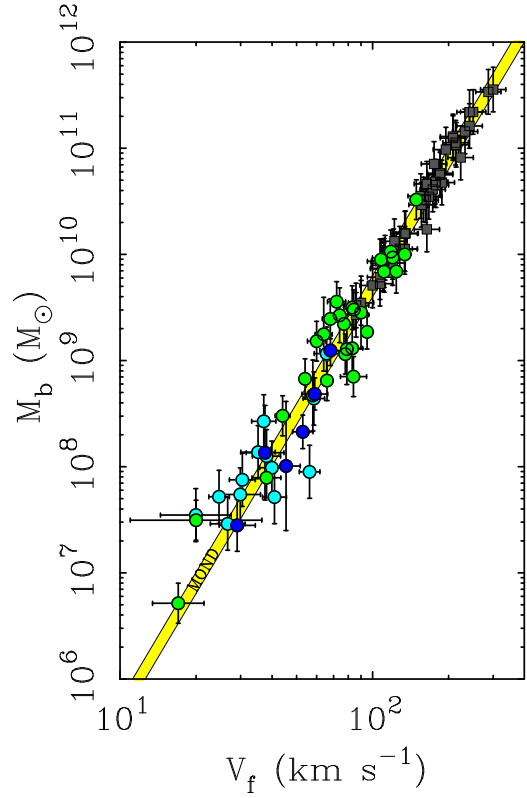}
\caption{The observed relation between the limiting rotational speed of galaxies, $V_f$, and their total luminous mass, $M_b$ (every point, with its indicated measurement errors, represents one galaxy). The (yellow) band shows the absolute MOND prediction (its width corresponds to the uncertainty in $\az$). Courtesy of Stacy McGaugh.}\label{fig4}
\end{center}
\end{figure}

\section{Enter dark matter}
Most scientists think that we should cleave to the standard laws of dynamics, despite the mass discrepancy. The disagreement between theory and observations is then explained away by assuming that galaxies and galactic systems are populated by large quantities of dark matter (DM): matter that is not yet visible directly by any of our instruments. This is matter whose exact composition is not even known yet. We know, however, for various reasons, that most of it cannot be made of any of the known forms of matter. This extra matter augments gravitational forces, and hence could bridge the mass discrepancy.
\par
According to this hypothesis, in every galaxy, the DM is present in the appropriate amounts and right distribution in space, to explain the observed gap between the calculated and observed velocities. The amount of DM required for this purpose, in a typical galaxy, can be tens of times larger than that of the standard, luminous matter. All in all, in the universe at large, one needs about five times more DM than luminous matter. Much of either is not locked in galaxies, but dispersed somehow in the inter-galactic space. In fact, most of the standard, potentially luminous, matter, believed to exist in the universe, is itself not really luminous, but lurks in hiding somewhere.
While this still unseen lion share of standard matter is also DM of a sort, it cannot constitute THE DM, since its total amount falls much shorter of what is needed, and also, its properties are not what is required from the DM.
\par
Another puzzle has emerged in recent years, which on the face of it is unrelated to that of the mass discrepancy: We have known for eighty odd years that our visible universe is expanding. This is indicated by the fact that the objects that stud the universe (e.g., galaxies) are observed to continually recede from each other. From this it is deduced that our part of the universe started its life in a most compact state, known as the ``big bang''.
\par
Until not so many years ago, we had all believed that the expansion of the universe is decelerating. The question remaining had been, then, whether the expansion will continue forever at an ever slowing rate, or will halt and reverse into contraction at some point in the future. This belief was based on preconceptions, not on observations: since the evolution of the universe is governed by its self-gravity, and since gravity can only attract, it follows that the expansion can only slow down. To most people's surprise, the expansion was found to accelerate (the recent Nobel prize went for this discovery). Here again, an expectation of the standard dynamics has been refuted. To explain this anomaly, many scientists have invoked a new ``dark'' entity, namely, ``dark energy'', whose best known property is, arguably, its name. It cannot be made of any kind of matter of standard properties, because such matter (dark or not) can only attract, and thus cause deceleration. Even though the puzzle of the ``dark energy'' is considered by many the most disturbing conundrum in physics today, I shall not enlarge here on this aspect of what has been called ``the preposterous universe''. This universe's material content is purported to be made up from about five percent ``luminous matter'', about twenty percent DM, and about seventy-five percent ``dark energy''.
\par
The working hypothesis of many physicists today is that the DM is made of elementary particles, whose existence has not been confirmed in the laboratory; it is only suggested by various theories that strive to extend the ``standard model'' of elementary particles. (In the past, there had been other favorite constituents of the putative DM, but they have been excluded in different ways.) If this working hypothesis is correct, then perhaps we can discover these DM particles directly on Earth: Our galaxy, as all others, is purportedly bathed in a halo of DM particles, and these are supposed to be everywhere, including Earth, our bodies, and our laboratories. These particles interact only very weakly with matter, and they easily penetrate most objects. It might then be possible to detect them by their rare interaction with the atoms of standard matter. There is a number of ongoing experiments to do this that employ specially designed detectors in underground laboratories. Much efforts have been put into such endeavors for over ten years now.
\section{Enter MOND}
In the early eighties of the last century, it occurred to me that the omnipresent mass discrepancy might be due to another cause, not to the presence of DM (the need for ``dark energy'' was not known then): One finds a discrepancy because one insists on applying the accepted laws of nature to systems like galaxies. Such insistence is not necessarily justified, because these laws have been distilled from studies of ``local'', small systems, such as laboratory aparati, or the solar system. It is possible that these laws, which do apply very accurately under such local conditions, do not apply to galactic systems. These latter are very different from the former in many regards; so, I thought that it might be possible to attribute considerable departures from the standard laws to such differences.
\par
Is it possible, then, to forgo DM, and formulate a new theory of dynamics by which the expected motions in galaxies will agree with the observations? In other words, is there such a theory by which the conventional, luminous matter, alone, provides enough gravity to fully counterbalance the observed motions, with no need for extra DM? Such a theory is also required to coincide with the standard dynamics under the conditions that prevail in the laboratory and the solar system, so as to retain the latter's successes there.
And, indeed, after trying several blind alleys, I found an initial formulation of such a theory, whose basic tenets are as follows:
\par
1. What distinguishes between galactic systems, on one hand, and the solar system, for example, on the other, is the characteristic accelerations found in these systems. The acceleration is the key property, on which hangs the difference between the standard dynamics and the new theory, which I dubbed MOND (for Modified Newtonian Dynamics). The acceleration measures the rate of change of velocities in the system; it is the direct measure of the gravitational forces in the system. The accelerations in galactic systems are many orders of magnitude smaller than those of planets in the solar system. For example, the acceleration of the sun, as a typical star, in its motion around our Milky-Way galaxy, is a hundred million times smaller than the acceleration of Earth as it goes around the sun, and a hundred billion times smaller than the acceleration of an object falling freely on Earth.
\par
2. MOND introduces into physics a new constant of nature, with the dimensions of acceleration, $\az$. This constant plays different roles in the new dynamics. First, $\az$ marks the boundary of the domain of validity of the standard dynamics (Newton's and Einstein's): in systems in which the accelerations are much larger than $\az$, such as the solar system, standard dynamics is a very good approximation to MOND (increasingly so, the higher the acceleration). In contradistinction, MOND departs considerably from the standard dynamics when the accelerations are smaller than $\az$, as in large regions of galaxies, for example. In fact, $\az$ disappears from the description of physics at high accelerations, while for physics at low accelerations, $\az$ appears in full force. And so, the stamp of $\az$ is predicted to appear--and is indeed seen--in many disparate galactic phenomena and regularities. These ubiquitous appearance of $\az$ in seemingly unrelated contexts, does not have an explanation without the unifying framework of MOND.
\par
Such roles of $\az$ are reminiscent of the different roles that Planck's constant $h$ plays in quantum theory, or the speed of light, $c$, plays in relativity. Like $\az$, they each mark a borderline, beyond which one has to use the modified dynamics they each represent.
And, like $\az$, they each appear in a variety of phenomena and regularities that come into play beyond this borderline; phenomena that are connected together only by the improved dynamics.
\par
In light of this, I have gained confidence, and drawn comfort, from the fact that laws of nature that had been accepted for centuries (Newton's laws) have already been subject to far-reaching amendments, when they failed to describe phenomena outside the realm in which they had been crystalized and tested before. Such an observation, in itself, must not instigate us to embrace freely various ``revolutionary'' ideas that are propounded all too often. But when there are cogent reasons for doing so, one can draw encouragements from such precedents.

\section{MOND predictions and their observational tests}
How, then, does MOND differ from the Newtonian dynamics when the characteristic accelerations in a system are smaller than $\az$?
A detailed description of the theory would be out of place here, but I'll draw a caricature that encapsulates its salient properties.
According to the Newtonian dynamics, a small object--a planet for example--at a distance $R$ from a large mass, $M$--the sun, for example--is subject to an acceleration, $a$, that is proportional to $M$, and is inversely proportional to the square of the distance--we write this as $a=GM/R^2$. MOND decrees that if the acceleration obtained from this relation is much larger than $\az$, then this relation still  holds to a good accuracy (the larger the acceleration the better the accuracy).
In other words, MOND and Newtonian dynamics agree in their predictions when the calculated acceleration is much larger than $\az$.
However, if the acceleration is much smaller than $\az$ we have to use another calculation for the acceleration felt by the body: Now the acceleration dictated by MOND is proportional to the square root of the mass, and inversely proportional to the distance itself, not its square. We then write $a=\sqrt{\az G}\sqrt{M}/R$. In between the two extreme limits there is a gradual transition between the two domains, described by the detailed theory. A schematic depiction of the Newtonian and the MOND laws is shown in Figure 1, for two values of the mass $M$.
\par
The main point to be gleaned from this succinct description is this: In the domain of small accelerations, the acceleration predicted by MOND is larger than that predicted by Newtonian dynamics, and the smaller the acceleration the larger the gap between the predictions. This fact stands in the basis of the MOND explanation of the mass discrepancy: the luminous mass alone subject bodies in its vicinity to accelerations that are larger--sometimes much larger--than those calculated from Newtonian dynamics. Gravity is thus enhanced (over its Newtonian strength) by the theory itself, with no need for extra (dark) matter. And so MOND bridges the mass gap.
\par
This simplistic statement is not the end of it, of course. The relevant observations pertain to hundreds of systems of very different attributes, such as masses, sizes, and shapes. Furthermore, each such system, in itself, is a cosmos in miniature that presents to us a complex picture of mass discrepancies: Not just one number that represents some global mass discrepancy for the system, but some distribution of mass discrepancies in the system. In other words, what we observe is not just the total luminous mass of the system, but its distribution, and what we calculate is not just the total mass required in the system, but its distribution; and the distributions disagree with each other in different ways at different locations in the system. So, a theory like MOND has to deal with, and account for, a wide variety of systems, and explain each of them in detail, without DM. We have no leeway in this game, apart for some remaining uncertainties in the astronomical observations themselves.
\par
Contrary to this very restrictive game for MOND, the DM hypothesis leaves us much freedom when we come to account for the mass discrepancy in a given system. There is no a priori commitment, no prediction, of how much DM there should be in a given galaxy, and of how it should be distributed. By this hypothesis we simply assume that what is needed is what is there. In light of this, the performance of MOND in accounting for galaxy dynamics are so impressive, especially that when MOND was propounded we had only a very small fraction of the observations we now have, and which MOND had predicted.
\par
MOND predictions, listed in the first papers, published in 1983, were later vindicated one by one (with one fly in the ointment, to be discussed below), including a detailed account of the velocity distribution in hundreds of galaxies. Also confirmed were universal laws predicted by MOND that galaxy properties should obey: laws that are partly analog to the laws of planetary motions, but which differ from Kepler's laws (which follow from Newtonian dynamics).
\par
Take, for instance, the crucial analysis of so called ``rotation curves'' of galaxies, arguably, the most clear-cut evidence for the mass discrepancy. A large fraction of the galaxies in the universe today are so called ``disc galaxies''--galaxies whose main component is a flat, thin, circular disc, made of gas and stars moving in circles of different radii around the galaxy (see Figure 2). The rotation curve describes the rotational speed of the stars and gas as a function of their distance from the center of the galaxy.
To analyze this facet of galaxy dynamics, one first measures the observed distribution of luminous matter in a galaxy; from it, one calculates the galaxy's gravitational field according to Newton's laws. From this one calculates the expected rotational speed at each radius (by equating the gravitational force to the centrifugal force at each radius--the two have to balance). This gives us the rotation curve of the particular galaxy as it is predicted by standard dynamics.
In Figure 3 we see, as an example, a rotation curve calculated in this way for a galaxy named NGC 1560 from Newton's laws assuming the presence of only the luminous matter (in blue). The figure also shows the rotation curve that was actually measured. We see that it differs greatly from what is predicted. The discrepancy between the velocities become as high as a factor of 2.5, which bespeaks a mass discrepancy of about a factor of 6 (masses go as the square of the velocities). In the framework of the DM paradigm, this means that within the observed region we need about six times more mass in the galaxy than is actually observed as luminous matter (from models, the discrepancy even increases beyond the radii depicted in the figure).
\par
In distinction, the green line shows the prediction of MOND for this galaxy, again, assuming the presence of only the observed, luminous matter. This is an absolute and unavoidable prediction, with hardly any leeway, except for some uncertainties in the astronomical parameter that were used (for example, the exact distance to the galaxy). The figure speaks for itself, as we see that MOND predicts even the finer variations in the measured curve. This picture repeats itself in over a hundred galaxies.
\par
As already mentioned, beyond the predictions concerning individual galaxies, such as the prediction of rotation curves, MOND predicts general laws of galactic dynamics. For example, MOND predicted that for every galaxy, the rotational speed should become independent of the orbital radius for large radii, as indeed was found to be the case (as we see in Figure 3 happens for NGC 1560). Another such predicted law is that the total luminous mass in a galaxy should be proportional to the fourth power of this asymptotically constant rotational speed. Kepler's laws, which rest on Newtonian dynamics, predict other behaviors: that the rotational speed should decline indefinitely with increasing radius, and that the mass should be proportional to the second power of the speed. This indeed happens with planetary motions in the solar system, but is in stark contrast to what we see in galaxies.
\par
This last-mentioned prediction of MOND has received a particularly acute confirmation in the work of the American astronomer Stacy McGaugh from the University of Maryland published last year. An example of such a test of the MOND prediction is shown if Figure 4.
\par
Regarding the historical development of the MOND paradigm, the first version of MOND was rather primitive in some regards, and its ensuing development is an interesting saga, for which I have little room here. Over the years, MOND was improved in different ways, but all its versions to date have been based on the same basic tenets, from which alone follow the salient predictions alluded to above. For this reason, practically all existing versions share these salient predictions, but they do differ among themselves in the predictions of more subtle phenomena.
\par
For example, the early versions described only non-relativistic phenomena. But over the years relativistic versions have been worked out (taking into account some  principles that we think should be retained from Einstein's relativity) with successive improvements. One important landmark was the formulation of a complete, relativistic MOND theory by Jacob Bekenstein from the Hebrew University, building on ideas by Robert Sanders from the University of Groningen. Recently, I proposed a relativistic MOND theory based on other principles (see below). Over the years, tens of researchers all over the world, whose deserving work cannot be described here, have contributed to the development of MOND. To date, more than five hundred scientific papers about MOND have been published.

\section{Deeper ramifications}
Quantum theory and the theory of relativity started their way as ``phenomenological'' theories: Their departure from the ``classical'' theory stemmed from the need to explain phenomena that remained unexplained by this theory. It has turned out, in the end, that beside being a better description of nature, these theories have ushered in new and far reaching concepts, concepts with no place in the older, ``classical'' world view. Quantum theory has begot the notion of discreteness of the possible states of a physical system (for example, the energy levels of an atom), and the uncertainty of the results of measurements, as a built-in principle.
The theory of relativity introduced the relativity of length and time, and the notion that the geometry of space-time varies according to circumstances, and is not a rigid, given arena for physics.
\par
MOND is, today, in a stage where it mainly answers a need to explain phenomena. But there are strong indications that our understanding of the theory now at hand is only the tip of an iceberg, and that MOND's foundations also rest on deeper strata. One very interesting clue is provided by the numerical value of $\az$. As I noticed as soon as MOND took shape, this constant--which can be measured in different ways, based on the various phenomena in which it plays a role--is close in its value to some acceleration parameters that are significant in the context of cosmology. For example, $\az$ is near in value to the acceleration that characterizes the accelerated expansion of the universe (alluded to above, and whose standard explanation is dark energy). It is also near in value to the ratio of the speed of light and the age of the universe since the big bang (in other words, an object accelerating with $\az$ from rest, will approach the speed of light in the lifetime of the universe).
\par
All this offers a possible clue to a tight connection between the ``mass discrepancy'' in galaxies, and the ``acceleration discrepancy'' of the universe: it points to the possibility that the conundrums of the DM and the dark energy have a common origin.
More generally, this coincidence points to the possibility that the global state of the universe enters and affects the dynamics in small systems, such as galaxies. Such a connection has no room in the physics known to us. Another clue to such a connection emerges from the existence of symmetries that are common to certain versions of MOND and to the universe at large.
\par
Another tantalizing pointer, of possible far-reaching implications, it raised by a relativistic formulation of MOND that I proposed recently, known as ``bimetric MOND gravity (BIMOND)''. According to this theory, we, and all that is within our ken, live, in some sense, in one of two parallel worlds. (Parallel worlds have been invoked in contexts such as the interpretation of quantum physics, and a multi-universe picture; but here the meaning is quite different.) When there is complete symmetry between the two worlds (the same matter content and distribution, the same interactions between matter constituents, the same starting condition) general relativity applies exactly (with a ``cosmological constant'' component). The departure from general relativity expressed in MOND, stems from the interaction between the two worlds, which arise locally as a result of deviations from the symmetry (e.g., access of matter in one world relative to the other).

\section{A period of struggle}
The struggle between the two paradigms for the hegemony is, of course, mainly a matter for physicists to deal with and settle. However, it also opens a window for edifying observations, to those interested in the history of science, and in the factors underlying the development of scientific theories.
\par
It is interesting to follow, for example, the different stages of rejection-acceptance of MOND by the community, beginning with an almost total disregard, to its status today as the rebellious contender, younger sister to the mainstream paradigm.
\par
It is also interesting to understand how such two contrary paradigms are each supported by first-rate scientists. After all, it appears that the same facts are lain before all of them, and the same arguments are known to all of them, and it is common knowledge that the scientific way is all based on cold and rational weighing of the facts. How then do different scientists arrive at so different conclusions from the same given? The main reason that such a state of things is possible is that the two paradigms have successes, but they also each have weaknesses, as each of the two performs better for one class of systems and less for others.
\par
For instance, MOND has unquestioned superiority in describing galactic phenomena. As I said above, MOND predicts, with great accuracy, the motions observed in individual galaxies, and also various general laws and regularities that characterize the population of galaxies--a collection of laws of galactic motions akin to Kepler's laws for planetary motions.
\par
In contradistinction, the DM paradigm faces difficulties in this quarter. Not only is its predictive power concerning individual galaxies much inferior to that of MOND, some of the predictions that it does make seem to be in direct conflict with observations, and many attempts are made to bridge such conflicts by postulating different complicated scenarios, or even by inventing new and sometimes additional types of DM.
\par
To me, the most poignant problem with the DM paradigm is that it does not predict, and is fundamentally incapable of predicting, for example, the rotation curves of individual galaxies, or the general rules I mentioned above. All these observed regularities, if we try to view them within the DM paradigm, would reflect tight relations between the luminous and dark matter in each galaxy. This, however, is untenable: According to this paradigm, a present-day galaxy is a product of a very complex, random, and violent history, involving collapse of gas clouds, their subsequent random mergers and swallowing up of other structures, part or whole, supernova explosions that expel matter from the forming galaxy, etc., etc.
The luminous and dark matter, because of their very different physical properties, are affected very differently by these processes; so, the connection between these two components in the final product, should, perforce, be very haphazard, contrary to the tight regularities observed. The DM paradigm will thus never be able to predict such a relation in an individual galaxy, whose construction history is unknowable. And, the very existence of tight regularities speaks against this scenario.
\par
Many DM advocates say they hope to one day be able to explain such regularities from the complex formation scenarios. I find such hopes similar to hoping that we will, one day, understand how Kepler's law, obeyed accurately by all planetary systems, follow from the complex (and not yet understood) processes of formation of such systems. A ludicrous hope (and not one entertained by anyone I know), since we know that Kepler's laws are a result of laws of physics (Newton's laws), and so must hold in any planetary system, regardless of the way it formed.
This is also the nature of the galactic laws according to MOND.
\par
On the other hand, MOND itself does not fully account for the mass discrepancy in clusters of galaxies: A Newtonian analysis of the observations points to a mass discrepancy of a factor of about ten
(the mass required is ten times the mass seen in these clusters). Analysis according to MOND reduces this discrepancy greatly, but still leaves a clear discrepancy of about a factor of two. Supporters of DM construe this as evidence that even MOND still requires some DM. The truth is, however, that this remaining gap can be explained without having to invoke a new kind of matter: even a small fraction of the still hiding normal matter, mentioned above, would suffice to explain the remaining cluster discrepancy. Be that as it may, DM advocates point to this as a weakness of MOND.
\par
In addition, MOND is not yet developed enough to present a full and accurate picture of the cosmological evolution of the universe as a whole, and the formation of structure in it. For this it is necessary
to have a fully acceptable relativistic version of MOND, and an understanding of the connection of MOND with cosmology. Despite much progress in this vein, leading to several candidate relativistic theories, there does remain much to be done in this direction.
\par
Add to these weaknesses of the two paradigms the factors stemming from human nature, such as preconceptions, interests, and variety in scientific tastes--which all have a major role in shaping the opinions of scientists--and we can understand how the two conflicting paradigms can live side by side, with the proponents of each pointing complacently to the successes of their favorite, and disapprovingly to the faults of its rival.
\par
Such a state of things is typical of interim periods, and is exemplified by many episodes in the history of science--the most famous perhaps being the century between Copernicus' death and Newton's birth, when the two coexisting rival world pictures were the Ptolemaic, geocentric, established view, and the Copernican, heliocentric, iconoclastic view.
\par
Another interesting phenomenon, also typical of such periods, is the appearance of hybrids of the two main competing paradigms, in attempts to enjoy the best of the two worlds. Such hybrids are propelled not only by a scientific impetus; psychologically, they also enable their supporters to adopt the successes of the new paradigm, while retaining their fidelity to the ``old slipper'' with which they feel comfortable.
\par
A well known example, of such hybrids, is Tycho Brahe's world picture, born in the days of the struggle between Earth and the sun on the center of the cosmos. This picture offers a compromise between the two contenders: It proposes that all the planets (excluding Earth) do revolve around the sun, but the sun itself, together with this entourage, revolves around the earth. This picture had shared some of the successes of the Copernican picture, yet kept Earth (and man) at the center of the universe. This paradigm also permitted a literal construction of the biblical ``Sun, stand still over Giveon'', and so was more palatable for the church.
\par
Similarly, there are today several suggestions to marry MOND with DM.
Historical precedents, together with my personal aversion to such hybrids--which together with the best of the two worlds, tend to carry also the worst--lead me to believe that such chimeras will fall by the wayside.

\section{In search of conclusive evidence}
I am often ask what I think will eventually decide the struggle.
Will it be an experimentum crucis, a crucial, single experiment or observation leaving no room for doubt? I do not foresee a decisive discovery that will fully convince one of the camps in the right of the other, barring a convincing discovery of DM by one of the present or future experiments that look for it on Earth.
\par
For example, as I already mentioned, in the beginning of last year
a result was published that is sharp and clear cut, that agrees precisely with a prediction of MOND made thirty years ago, and which is very difficult to understand in the context of the DM paradigm (let alone predicted by this paradigm). This finding is not inferior in its accuracy and ramifications to the measurement of the bending of light rays by the sun's gravity (in 1919), which agreed with the prediction of general relativity, and greatly contributed to its acceptance. Then too, it was possible, in principle, to explain the observations as resulting from some refractive, transparent matter around the sun, with just the right properties to account for the observed bending. (I am not aware of actual attempts at such a distorted explanation to try and save the conventional theory, although then, like today, doubts have been cast by detractors on the validity of the observations themselves.)
Indeed, in the MOND camp this result is perceived as an almost decisive evidence, but from the DM camp one hears mostly ``yes, but...''.
\par
And so, as we well know human nature, I suppose that the process of acceptance of MOND will be gradual, if indeed it will win eventually.
If more and more time passes by, with no direct detection of DM, we are bound to see a decline in the support for DM. In addition, if MOND continues to show successes, the number of its supporters will increase, at least from the ranks of the younger researchers who join the research circle with few preconceptions.
\par
In any event, even before the struggle has been decided, MOND has already proven itself an important and useful paradigm: It has lead to the identification of new phenomenological laws in the data forest, which only a scrutiny through MOND glasses could reveal.
For example, that the galactic mass discrepancy sets in at accelerations smaller than some critical value, and that below this value the discrepancy increases in inverse proportion to the acceleration; for example, the tight relation between the rotational speed at the outskirts of a galaxy, and the galaxy's total luminous mass. And especially, MOND highlighted the ubiquitous appearance of the constant $\az$ in a variety of seemingly unrelated galactic phenomena. The importance of $\az$ is a {\it fait accomplis}; it is here to stay, be the results of the struggle what they may.
\end{document}